# Poseidon: Non-server WEB Forms Off-line Processing System

VACLAV SKALA
Department of Computer Science and Engineering
University of West Bohemia, Faculty of Applied Sciences
Univerzitni 8, CZ 30614 Plzen
Czech Republic
http://www.VaclavSkala.eu

*Abstract:* - The proposed Poseidon system is based on e-mail services of filled forms instead of WEB server based services. This approach is convenient especially for small applications or small–medium companies. It is based on PDF forms that are available on a WEB page. PDF forms can be downloaded, off-line filled in, printed and finally sent by e mail for final processing. Data are actually stored in the local outbox waiting for a connection to a mail server. This follows an idea of the standard "paper" letter sending. Filled in data are actually sent when a user is on-line, therefore a user is "free" of being on-line when filling the forms. When the PDF form is processed on the recipient side, answer is sent back via e-mail as well. Typical application is e.g. in conference management systems, systems for submission to journals etc.
The great advantage of the PDF forms use is that they can be easily made or modified by a non-specialized administrative person easily.

*Key-Words:* - WEB forms processing, e-mail services, non-server based system, conference management system, small business system

## 1 Introduction

Majority of today's WEB based applications or systems are based on on-line WEB forms made in HTML etc. and processing using https protocol and WEB server services. This approach relies on a perfect connection. If the connection is disrupted or filling takes a long time due to some reasons data are usually lost and the form filling starts from the scratch.

The forms cannot be filled in off-line in principle, nor data stored. The on-line forms filling requirement is crucial in many cases, e.g. if forms are to be filled in on long distances trains or during a flight, connection cost is high especially if travelling abroad or if a satellite connection is used, in rural areas an on-line connection is not usually available.

There is also another crucial requirement – storing exact data before actual sending, not taking a copy of data delivered back from the WEB server as usual. Also a necessity to store data and information - even very private - on a server operated by some company (in many cases without relevant security certificates and responsibility for data leakage) and the need to pay for the disk storage memory used or reserved etc. Also data security or data lost is very important issue, as the WEB server system administrators do have an access to all data, in general, if data are not encrypted.

There is also a dependency on software used for WEB forms processing. When software updated, e.g. PHP, CGI etc., the system has to be re-checked or modified, sometimes significantly, that is very costly at the end and at the user's account.

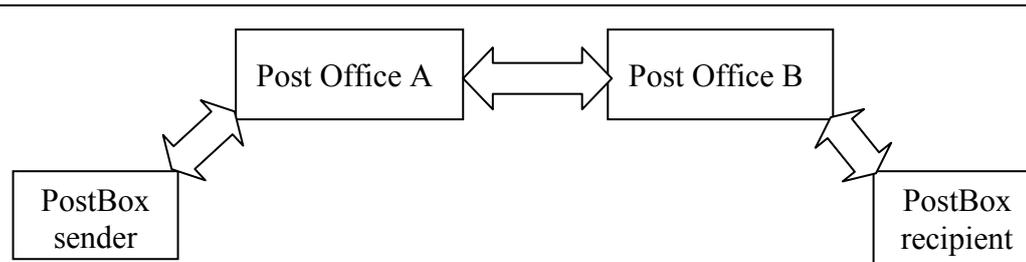

Fig.1: The Poseidon system architecture

As some WEB form based systems are quite complicated, is nearly impossible to check all possible cases and avoid some errors in principle.

Even more, in many cases the owner of the WEB based application is not informed in advance that something is going to be changed. There is also a severe problem with a detection which part of the application could not work properly, partially or in a total. In many cases, application owner or application users are informed on problems with a time delay, which might be a critical delay or a loss of a business contract etc.

Also this approach "forces" a user not to change the server provider due to the additional cost, i.e. reinstallation of the application, testing and error correction, if change is made.

In the following an experimental system based on e-mail services will be shortly introduced, see Fig.1.

It is primarily intended for small business and future mobile applications and for conference management system and for cases, where on-line connection is not 100% reliable or available from time to time only.

## 2  WEB On-line System Architecture

WEB forms are based primarily on the HTML forms generated at a server and processed by a server after filled by a user, send back to a server, where data are stored. This is a typical example of the Client-Server approach. Textual information and forms windows are mixed all together within HTML file and specialized HTML editors have to be used to produce the final code. In many cases the HTML editors are not mutually code compatible and produced code is not always 100% compatible with the W3C, especially if an older editor is used. It results to necessity of testing WEB pages in different browsers and their versions. Also testing on different platforms and operating systems is needed. This approach requires deeper knowledge of programming and HTML structure and lot of man-power to make it, etc.

The WEB forms have to be filled in when a user is on-line and connection is not interrupted. Data from the filled form are submitted to a server and processed by php script or by specialized programs, e.g. by a CGI script is used, etc.

All those techniques strictly require a user to be connected to a server all the time. This requirement is not fully acceptable especially for a mobile applications, e.g. if the forms are to be filled in a train passing through a tunnel or areas without a signal or on a plane board during a flight etc. If the connection is interrupted, data are lost. Therefore WEB forms use JavaScript and cookies very often to enable "re-login" without necessity to type username and password again.

When a connection is interrupted and cookies are allowed, last saved content can be retrieved, however data filled in after the last save operation are usually lost.

The Client-Server based applications have some *advantages* like simple maintenance or immediate response and data processing etc. However there are some *significant disadvantages* for small companies like:

- there is a necessity to store data and information - even very private data - on a server operated by a company and pay for the disk storage memory used or reserved etc.,
- there is a dependency on a software used for WEB forms processing on the server and when software updated, e.g. php etc., the system has to be "re-checked" or modified, sometimes significantly, that is very costly and time consuming at the end,
- time of connection during filling forms has to be paid, that might be very costly especially while travelling abroad or on a board of a plane etc.,
- Client-Server approach "forces" a user of the application not to change the provider due to the additional cost if change is made,
- necessity to have fixed public IP address if not-hosted server is used, but server is operated by a user,
- incompatibility of WEB browsers, i.e. MS IE, Google Chrome, Mozzila etc. and necessity to check how the WEB page is actually displayed,
- filled in HTML pages usually have a different shape when displayed and printed, in many cases and data inserted are usually not printed at all or a part of them are lost,
- there is a problem with a security due to JavaScript and cookies used and with attacks based on huge number of submissions.

Nearly all WEB pages are based on SOAP (Simple Object Access Protocol).

In the following we describe an alternative solution based on PDF forms and @mail based services, see Fig.2.

## 3  WEB Off-line System Architecture

An alternative approach to the WEB on-line filling is off-line filling using PDF forms or similar and submitting them after filling via e-mail services. This approach is "server-free" actually, as the server is needed only for forms to be downloaded by a user. The forms can also be sent directly to a user by an e-mail, if needed.

PDF forms stored locally can be filled in when a user is off-line and can be sent out at time the user connects to an internet. The filled in forms can be printed[1] for documentation as well. When data generated by PDF forms (PDF forms usually send filled in data in XML or similar formats as an attachment) are actually stored in the Outbox folder of the sender's @mail client and sent out by an e-mail server and waiting for a recipient's login and waiting for a download to recipient's Inbox folder (might be locally based if POP is used). When recipient log in, data are downloaded and processed. Data are generally stored locally and receipt of the message can be automatically generated and sent to sender of the form via e-mail.

When an e-mail client is connected all the data are saved to a local e-mail client's folder and the data at the e-mail server can be removed. This is a typical example of "Client-Client" application with a specific feature – clients need not to be on-line all the time as the e-mail server serves as queue of data to pre processed and for generated e-mails to be sent.

This approach has some *disadvantages*, of course, namely:
- data processing is not immediate, as the data processing is off-line,
- PDF forms have to be downloadable from WEB or sent directly to a user by e-mail,
- e-mail client at the users side is used to sent the e-mail server using *https* protocol, however it is not encrypted if the user does not specify that, but the XML file can be encrypted.

However, on the other hand there are *significant advantages* as:
- PDF files can be filled in and processed while a user is off-line,
- PDF forms can be printed to a PDF file in "1:1" formatting style with data filled in using "Print to PDF" printer's option,
- data are sent usually as XML file, i.e. the amount of transferred data is limited, that is important when a paid connection based with a limited data amount is used,
- the e-mail server serves actually as a buffer for data processing, virus removal, DoS based attacks etc.,
- all data are stored at the recipient's side where the client serves actually as a server and stores not only data sent, but also other data needed in the application,
- e-mail message itself can be protected if the mailing client at the user side enables that, e.g. using public key etc.,
- only a plain text can be sent via XML and therefore viruses infections are improbable.

As PDF forms are used for the WEB off-line architecture standard PDF form design applications can be used, e.g. Adobe PDF. Design of the PDF form is simple and even a non-programmer can make it easily and effectively.

It means that making a PDF form can be made by administrative people if they follow some basic simple rules. This decrease a cost and increase time flexibility as no programmer's work is needed.

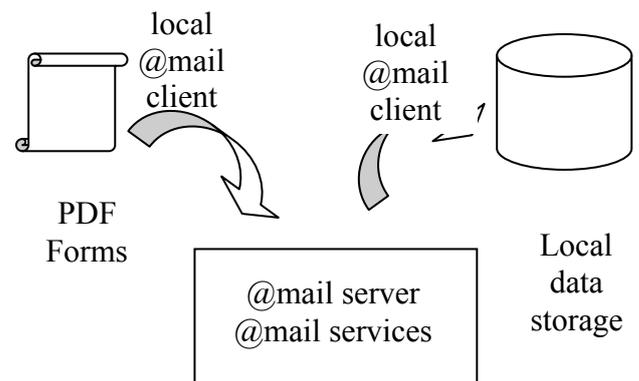

Fig.2.: Principle of the non-server approach

---

[1] Some PDF forms do not allow filled in to be printed directly. In this case a PDF printer, like PDF creator, has to be selected. It makes PDF file containing filled in data and can be used as a "hard copy" of the submitted information.

Fig.3.: Person registration for a conference application

## 4 Poseidon System

The Poseidon system is based on the WEB off-line architecture and it is intended for small and medium conferences management system and standard small office applications. All the PDF forms are available on "static" WEB pages or can be sent directly to potential users via email to hide them from WEB crawlers. Fig.2 presents an example of a person registration to a conference.

The Poseidon system application for conference management system contains several such pages, namely:

- person registration – a person registers name, institutional affiliation, country, his @mail and personal WEB page,
- paper registration – serves for a paper registration, i.e. title of the paper, authors and co-authors, abstract, keywords, address where the document is stored for download,
- reviewer registration,
- reviews submission,
- conference attendee registration,
- conference fee payment, etc.,
- conference attendee forms for VISA issuing.

PDF forms are easy to produce in general, if convenient PDF Form generator is used, and can be made by acknowledged administrative person in a conference or small business offices. It is also possible to add on additional scripts, e.g. JavaScript or FormCalc, for verification etc.

## 5 Technical Issues

As the Poseidon system relies on @mail based services, it is quite robust against attacks, as attacks will be actually against the mailing system itself, i.e. against the @mail services provider, which has very good defense system installed. It means that the "server" user does not need sophisticated security systems or a specialist handling attacks etc. The only case to be solved, that non-relevant e-mails have to be removed and multiple submission of the same filled form have to be correctly processed. It is quite simple, as in many cases just last re-submission is considered as correct.

As the e-mail client at the recipient's side can be set so it connects every $xx$ mins., communication between @mail server and recipient's @mail client does not introduce any significant communication load. The given approach also eliminates problem with number of connections, which could be a limitation factor in some cases.

## 6 Experimental Implementation

An experimental implementation for verification of the proposed approach uses Adobe PDF Forms producing XML file sent via e-mail, the standard mailer MS Outlook and CSV format for data processing, e.g. by MS Excel or Open Office, currently. Data are stored in the XLSX format Excel format and a layer for simplified SQL handling data will be shortly available.

All data are stored at the recipient's side together with other data needed for the specific office application/agenda. This brought quite significant simplification in interconnection with other data sources needed for the actual application, i.e. conference management system in our case.

The most difficult part is actually extraction of append files from the actual e-mail message on the recipient's side, as some e-mail clients do not have open interface for making that.

## 7 Experience

During first experiments one additional *advantage* of the proposed approach was found as it is easy to interconnect other data sets on the client side as proprietary software can be run directly on the recipient's side. The recipient's e-mail recipient's client can be active in selected time or repeatedly active on a given schedule.

On the other hand users should be clearly warned that the communication is not on-line, i.e. answers are not delivered and processed immediately. From the psychological point of view it forces a data sender to think more what he is actually sending, which is in many cases very positive at the end. The system was partially verified using e-mail hosted on a free mailing system

Fig.4.: Active PDF forms on a NOKIA LUMIA

It is necessary to point out that some PDF Form editors use verification of the content inserted to the form using validation scripts. In some cases, when a user uses "Cut-&-Paste" operations especially from WEB and MS Word documents, invisible characters are inserted, validation script does not accept such data and forms might not be sent out. A user should be warned in this case, if validation scripts are used.

## 8 Mobile Devices

The Poseidon system has one *significant disadvantage* – active PDF forms in Acrobat Reader on some Mobile platform are not currently supported on some mobile devices according to experiments made on IPHONES 5S with IOS &.x operating system and NOKIA LUMIA 920 with MS Windows 8 operating system. It means that forms cannot be filled in and sent via @mail.

## 9 Conclusion

The proposed approach used in the Poseidon system is simple and easy to use. As it is based on e-mail based services it is secure as

- all the data have specific data type, actually plain ASCII code,
- simple fake submission detection (sender & reply to addresses must be same, etc.),
- all data stored in local data storage (if the system is not run on the server directly) that means high data security – no access to other person including company administrators,
- possibility of digitally signed messages,
- user can easily modify the forms content easily,
- generally free of license software usage
- no WEB server dependency,
- no problems with software updates on a server side,
- no necessity to be on-line for processing data submitted,
- no necessity to be on-line while filling the forms.

This approach has some disadvantages, namely: no immediate response as the mail client is not on-line (note that it might be your personal mailing client in your mobile or notebook).

However the presented approach and the Poseidon system seem to be flexible for applications in small business applications due to its flexibility. In future work the principle of the Poseidon system will be generalized for small business and mobile applications.

## Acknowledgment
The author thanks to students and colleagues at the University of West Bohemia, Plzen for their critical comments and recommendations. Thanks belong to anonymous reviewers for comments and hints that helped to improve this manuscript significantly. This research was supported by project LH12181.

*References:*
[1] Bauknecht,K., Pröll,B., Werthner,H.(Ed.): E-Commerce and Web Technologies, LNCS 3590, 6th Int. Conference, EC-Web 2005, Proceedings, ISBN: 978-3-540-28467-3, 2005
[2] Bouchiha,D., Malki,M., Djaa,D., Alghamdi,A., Alnafjan,K.: Semantic Annotation of Web Services: A Comparative Study, in » Software Engineering, Artificial Intelligence, Networking and Parallel/Distributed Computing, Studies in Computational Intelligence, Vol.492, pp.87-100, Springer Verlag, 2013
[3] Covic,Z., Szedmina, L.: Security of web forms, 5th International Symposium on Intelligent Systems and Informatics, IEEE SISY 2007, pp.197-200, 2007
[4] Nolan,D., Lang,D.L.: XML and Web Technologies for Data Sciences with R, ISBN: 978-1-4614-7899-7, Springer, 2014
[5] Saha,T.K., Ul-Ambia,A.: Code Generation Tools for Automated Server-side, Int. Journal of Computer Science and Management Research, ISSN 2278-733X, Vol. 2, No.1, 2013
[6] Suddul,G., Nissanke,N., Mohamudally,N.: An Effective Approach to Parse SOAP Messages on Mobile Clients, in Mobile Web Information Systems - MobiWIS 2013 proceedings, pp.9-21, Springer Verlag, 2013
[7] Tao,Z.: Detection and Service Security Mechanism of XML Injection Attacks, in Yang,Y., Ma,M., Liu,B. (Eds.): ICICA 2013 conference proceedings, pp. 67-75, Springer Verlag, 2013
[8] Vogels,W.: Web Services Are Not Distributed Objects, IEEE Internet Computing, ISSN 1089-7801, Vol.5-6, pp.59-66, 2003
[9] Varnagar,R., Madhak,N.N., Kodinariya,T.M., Rathod,J.N.: Web Usage Mining: A Review on Process, Methods and Techniques, IEEE International Conference on Information Communication and Embedded Systems (ICICES) 2013, vol., No.2, pp.40-46, Feb. 2013